\documentclass[runningheads,a4paper]{llncs}
\pdfoutput=1
\usepackage{amsmath,amssymb}
\usepackage{epsfig}
\usepackage{listings}
\usepackage{multirow}

\newcommand{\QLname}{SimSeQL}

\begin{document}

\mainmatter

\title{Query Language for Complex Similarity Queries}

\author{%
Petra Budikova \and
Michal Batko \and
Pavel Zezula}

\institute{Masaryk University, Brno, Czech Republic}

\maketitle

\begin{abstract}
For complex data types such as multimedia, traditional data management 
methods are not suitable. Instead of attribute matching approaches, access 
methods based on object similarity are becoming popular.
Recently, this resulted in an intensive research of indexing and
searching methods for the similarity-based retrieval. Nowadays,
many efficient methods are already available, but using them to
build an actual search system still requires specialists that tune the
methods and build the system manually. Several attempts have already 
been made to provide a more convenient high-level interface in a form of query
languages for such systems, but these are limited to support 
only basic similarity queries. In this paper, we propose a new language that
allows to formulate content-based queries in a flexible way, taking
into account the functionality offered by a particular search
engine in use. To ensure this, the language is based on a general
data model with an abstract set of operations. Consequently, the
language supports various advanced query operations such as
similarity joins, reverse nearest neighbor queries, or distinct kNN
queries, as well as multi-object and multi-modal queries. The
language is primarily designed to be used with the MESSIF framework
for content-based searching but can be employed by other retrieval
systems as well.   

\end{abstract}

\section{Introduction}

Information has always been a valuable article but it has always
been difficult to obtain. These days, we have an unprecedented
advantage of having huge and rich data collections at our
fingertips. On the other hand, we still need more efficient tools
for data management to be able to locate the desired information
in the vast amounts of resources.
With the emergence of complex data types such as multimedia,
traditional retrieval methods based on attribute matching are no
longer satisfactory. Therefore, a new approach to searching has
been proposed, exploiting the concept of similarity between complex
objects. In recent years, we have witnessed intensive research in
the field of indexing methods and search algorithms for
similarity-based retrieval. As a result, state-of-the-art search
systems already support quite complex similarity queries with a
number of features that can be adjusted according to individual
user's preferences.

To communicate with such a system, it is either possible to employ
low-level programming tools, or a higher-level communication interface
that shields users from the implementation details employed by the
particular search engine. As the low-level tools can only be used by
a limited number of specialists, the high-level interface becomes a 
necessity when common users shall be allowed to issue advanced 
queries or adjust the parameters of the retrieval process. 
In this paper, we are proposing such high-level interface in a form
of a structured query language that allows users to issue actual
queries over complex data.

The motivation to study query languages arose from the development
of our own framework for similarity searching called
MESSIF~\cite{messif07lncs}. The system currently supports a wide
spectrum of retrieval algorithms and is used to support several
multimedia search applications, such as large-scale image search,
automatic image annotation, or gait recognition. So far, users are
allowed only to select the query object via a graphical interface,
and the choice of the actual search methods as well as its
parameters and other settings are hard-coded into the system. To
improve the usability of our systems, we decided to provide the
framework with a query language that would allow advanced users to
express their preferences without having to deal with the technical
details. After a thorough study of existing solutions we came to a
conclusion that none of them suits our specific needs. Therefore, we
decided to propose a new language based on and extending the
existing ones. At the same time, it was our desire to design
the language in such a way that it could be also used by other
systems. 

Consequently, we present here an SQL-based query language which can
be used to formulate a wide range of similarity queries, as we
demonstrate on examples from various application domains. Building
on a thorough analysis of previous studies and our long-time
experience with both theory and practice of similarity search
systems, we have proposed its structure so that it supports all
fundamental query types and can be easily extended. The language
can be used by programmers or advanced users to issue queries
in a standard declarative way, shielding them from the execution
details. For less advanced users, we expect the language to be
wrapped-up into a visual interface. The language is designed in a
general way as to allow flexibility and extensibility. 

The paper is further organized as follows. First, we review the
related work in Section 2. In the following section, we analyze the
requirements for a multimedia query language, taking into account
current trends in information retrieval research, lessons learned
from other language proposals, and the functionality of the MESSIF
framework. Next, we discuss the fundamental design decisions that determined
the overall structure of the language in Section 4. Section 5 introduces
both the theoretical model of the language and its syntax and semantics. 
Section 6 presents several real-world queries over
multimedia data, formulated in our language.
Finally, we outline the future work in Section 7.

\section{Related Work}
\label{sec:related-work}
 
The problem of defining a formal apparatus for similarity queries
has been recognized and studied by the data processing community
for more than two decades, with various research groups working on
different aspects of the problem. Some of these studies focus on
the underlying algebra, others deal with the query language syntax.
Query languages can be further classified as SQL-based, XML-based,
and others with a less common syntax. We shall briefly
survey all these research directions.

Similarity algebra as a tool for theoretical modeling and
transformations of similarity queries was first introduced
in~\cite{Subrahmanian98}. The authors define general abstractions
for objects and similarity measures, present basic algebra
operations and discuss their properties. Later works add new
similarity operations~\cite{Atnafu01} or study the integration of
similarity-based querying into established data models, e.g.
relational model~\cite{Belohlavek10}. While these studies provide a
valuable insight into the principles of similarity searching, the
algebraic operations used to express the queries are not meant to
be employed by users during a search session.

The majority of the early proposals for practical query languages
are based on SQL or its object-oriented alternative,
OQL~\cite{Cattell93}. Paper \cite{MOQL} describes MOQL, a
multimedia query language based on OQL which supports spatial,
temporal and containment predicates for searching in image or
video. However, similarity-based searching is not supported in
MOQL. The authors of \cite{Amato97} introduce new operators {\tt
sim} and {\tt match} for object similarity and concept-object
relevance, respectively. However, it is not possible to limit the
similarity or define the way it is evaluated. In \cite{GaoWWP04}, a
more flexible similarity operator for near and nearest neighbors is
provided but it still does not allow to choose the similarity
measure.

Much more mature extensions of relational DBMS and SQL are presented 
in~\cite{BarioniRTT05,BarioniRTJ09,POSTGRESQL-IE}. The concept of~\cite{BarioniRTT05,BarioniRTJ09} 
enables to integrate 
similarity queries into SQL, using new data types with associated similarity measures and extended 
functionality of the select command. The authors also describe the processing of such extended SQL 
and discuss optimization issues. Even though the proposed SQL extension is less flexible than we need, 
the presented concept is sound and elaborate. The study~\cite{POSTGRESQL-IE} only deals with image 
retrieval but also presents an extension of the PostgreSQL database management system that enables 
to define feature extractors, create access methods and query objects by similarity. This solution 
is less complex than the previous one but on the other hand, it allows users to adjust the weights 
of individual features for the evaluation of similarity.

Recently, we could also witness interest in XML-based languages for
similarity searching. In particular, the MPEG committee has
initiated a call for proposal for MPEG Query Format (MPQF). The
objective is to enable easier and interoperable access to
multimedia data across search engines and repositories. As
described in~\cite{DollerMPQF08}, the MPQF consists of three
fundamental parts -- input query type, output query type, and query
management tools. The format supports various query types (by
example, by keywords, etc.), spatial-temporal queries and queries
based on user preferences.  It also supports result formatting and
foresees service discovery functionality. From among various
proposals we may highlight~\cite{Tsinaraki07} which presents an
MPEG-7 query language that also allows to query ontologies
described in OWL syntax.

Last of all, let us mention several efforts to create easy-to-use
query tools that are not based on either XML or SQL. The authors
of~\cite{Schmitt05} propose to issue queries via filling table
skeletons and issuing weights for individual clauses, with the
complex queries being realized by specifying a (visual) condition
tree. In~\cite{Pein08}, a simple language based on Lucene query
syntax is proposed. Finally, \cite{Town01} describes a rich
ontological query language that works with structured English
sentences but requires advanced image segmentation and domain
knowledge.

\section{Analysis of Requirements}

Our objective, as mentioned previously, is to create a query
language that can be used to define advanced queries over
multimedia or other complex data types. The language will be
implemented on top of the MESSIF software, which is a framework for
creating similarity-based retrieval systems. Naturally, we also
want the language to be general and extensible, so that it can be
employed in a wide range of applications. To achieve this, we first
need to define the desired functionality of such a language.

In this section, we study the following three issues that are
closely related to the language design: (1) the current trends in
multimedia information retrieval, which reveal the advanced
features that should be supported by the language; (2) existing
query languages and their philosophies, so that we can profit on
previous work; and (3) the MESSIF framework architecture, which
should be compatible with the language. After a thorough analysis
of these sources we compose a structured list of requirements. 

\subsection{Current Trends in Multimedia Information Retrieval}

Contemporary science distinguishes two basic approaches to searching in digital data -- the attribute-based searching~\cite{DBbook} that is used in the traditional DBMS, and the similarity-based 
retrieval~\cite{Zezula2006}. In the first case, queries are defined by a set of strict conditions that 
are applied on attributes of data objects and the qualifying objects are returned. In similarity-based
retrieval, queries are usually defined by an example object and objects most similar to it form the 
response. The similarity can be described by a distance function, the smallest distance representing the 
best similarity. Alternatively, the similarity can be expressed as
a score where higher scores denote more similar objects.
Since these two approaches are interchangeable, we will use
the distance terminology from now on.
The most commonly used similarity queries are the {\em k-nearest neighbors query} (kNN) and the {\em 
range query}; the first restricts the number {\em k} of the most similar objects to be retrieved, the 
second limits the search by the maximum distance of a qualifying
object. However, there exist a number of other query 
types, such as various sorts of {\em similarity join}, {\em reverse nearest neighbor query}, 
{\em skyline query}, {\em distinct kNN query}, etc.~\cite{Zezula2006} 

In order to enable efficient retrieval, any search method needs to
be backed by a suitable data 
management structure. The indices used for attribute-based and similarity-based retrieval are 
substantially different. The traditional solutions used in relational databases employ index trees 
that organize data using the total ordering property of individual data domains. In content-based 
searching, the data domains frequently do not have this property and the objects need to be organized 
with respect to mutual distances only. In consequence, the indices for similarity searching usually 
cannot support attribute-based queries and vice-versa. Therefore, these two approaches to searching 
need to be considered as independent and complementary.

The attribute-based approach is long-established and well-tuned but
it is known to be unsuitable for complex data such as multimedia,
since exact match queries can only find binary-identical content
and the metadata is often not expressive enough or not available at
all. Similarity-based methods enable to search the complex data in a
more natural way but they also have some limitations. The retrieval
methods typically employ low-level content descriptors, such as
color histograms in case of an image, which are far from human
understanding of the object. The discrepancy between the object
descriptor level and human-perceived semantic level is often
denoted as the {\em semantic gap} problem~\cite{Smeul00}, which is one of
the major challenges in multimedia retrieval nowadays. 

Recent works~\cite{Datta08,Jain10} suggest that promising results can be achieved by combining the two 
above-mentioned approaches together. Attribute-based and simila-rity-based retrieval are orthogonal 
to each other and their composition can cover both the content of the object and its semantics. Let us consider the following query: {\em Retrieve all information about a flower similar to this photo, which 
grows in the Alps and blossoms in spring}, which includes both an example data object and strict 
conditions on some of its metadata.  Such query can be evaluated in several ways -- the system can 
first execute a content-based query and then filter the results, or start with the attribute restrictions, 
or evaluate several separate sub-queries and combine their results.
Each of these execution plans may be suitable in different
situations. Therefore, an advanced query interface should allow
users to define how a combined query should be processed. Support
for both types of searching, the various query types and their
combinations needs to be part of a query language.
 
An important issue connected with complex data searching is the formulation of a search task. Frequently 
it is not possible to define the query in a precise way. Instead, a user may describe the desired result 
by several conditions together with a specification of their importance. Typically, the individual 
conditions may have weights assigned to them. With the query-by-example paradigm, it is also often 
difficult to obtain a really representative query object. To overcome this, it is necessary to support 
queries with multiple examples as well as iterative searching with relevance feedback. Moreover, it is 
desirable to allow users to alter the definition of object similarity, as this may vary for different 
people and situations. There may also be additional parameters of the search process that users want 
to control, such as the cost/precision ratio for large data processing. Apart from including the features 
mentioned so far, which are perceived as necessary in most studies, the language should allow easy 
integration of other functionality that may be needed in applications, such as new query types or search 
algorithms.

\subsection{State-of-the-art Query Languages}

In this section, we analyze the main requirements and functionality
that can be encountered in various works on query languages
surveyed in the Related work. Some of the requests were formulated
explicitly, especially in the MPEG Query Format, others were picked
from the design of the individual languages. The identified
features fall into the following categories:

\begin{itemize}   
\item Support for similarity queries: Many of the existing studies focus on introducing query language 
primitives for basic similarity queries -- the kNN query, range
query, and several types of similarity 
joins are mostly considered. Typically, a special primitive is designed for each query type. Different 
keywords are introduced in the individual languages.
\item Integration of attribute-based and similarity queries: The need for combining the two approaches 
to searching is recognized in various proposals. Most often, the integration is performed by incorporating 
the similarity search algorithms into a relational DBMS. 
\item Support for spacio-temporal queries: Some of the languages, including the MPEG Query Format, give 
special attention to queries concerning spatial and temporal characteristics of a multimedia object. 
In~\cite{MOQL}, a set of operators is designed to support this type of queries.
\item Adjustability of searching: There are a number of parameters of the search process that users may 
want to adjust. The ones that are most frequently supported in existing proposals are the weighting of 
search conditions and the definition of a distance function. 
\item Optimization issues: Optimization strategies strive to maximize the efficiency of query processing 
by evaluating the individual search operations in the most suitable order. To allow optimization, it is 
necessary to understand the priority of operators, their evaluation costs and the equivalences of expressions.
Several optimization rules can be found in~\cite{BarioniRTT05} concerning kNN, range and join query 
operators. As observed in~\cite{BarioniRTJ09}, the more specialized operators we introduce, the more 
precise optimization rules can be defined and vice versa.
\item Output formatting: In relational DBMS, output formatting options are limited to the choice of 
attributes and the ordering of tuples. Proposals of~\cite{DollerMPQF08,MOQL} expand this with the result 
paging option and result layout specification, respectively. 
\item Service discovery: As the MPEG Query Format aims at creating a uniform access interface to various 
search services, it also provides functionality for service discovery. In particular, it allows to ask 
the search engine for supported query types, metadata, media types and expressions, and to inquire about system usage conditions.
\end{itemize}

\subsection{MESSIF Architecture}

Metric Similarity Search Implementation Framework
(MESSIF)~\cite{messif07lncs} is a
Java-based object-oriented library that eases the task of
implementing metric similarity search systems. It provides various
modules that are commonly needed by search engines such as memory
and disk storage backends, network communication tools, statistics
gathering and logging tools, and so on.

The framework also offers an extensible way of defining data types
and their associated metric similarity functions and provides
implementations of several common data types and their typical
distances, e.g. vectors with $L_p$ metrics. MESSIF-enabled indexing
methods that utilize only the generic properties of the similarity
functions are then applicable to any such data type.

Finally, the framework offers generic hierarchy of data
manipulation and querying operations. Typical engine operations
such as insertion or range and kNN queries are of course
implemented as well as various other queries including the
similarity join or combined and multi-object queries. The
definition of new operations is also possible and easy. When
executing an operation, the framework automatically chooses the
evaluation plan either by using an index structure that is able to
answer the given query efficiently or by a sequential scan if there
are no usable indexes. Moreover, the precise or approximate
evaluation strategy (typically early-termination or pruning
relaxation) can be specified for most queries and taken into
consideration by the framework while evaluating the queries.

Overall, the framework offers functionality of specifying the data
type, the metric function, the type of similarity query and its
evaluation strategy by means of programming API. By defining the
query language we would allow to utilize this functionality without
the need for actual Java coding.

\subsection{Requirements Summary}

Obviously, there are a number of features that need to be
considered in the design of a query language for advanced
multimedia searching. Unfortunately, not all of them can be fully
satisfied as it is hardly possible to provide a language that is
general, extensible, and simple at the same time. In order to gain
more insight into the problem, we try to identify the main
involved parties and summarize their concerns:

\begin{itemize}
\item "User interest": The most obvious party is the end-users, who are often mainly interested in easy 
usability of the language. For a typical non-expert user, we should create a tool that allows to 
formulate any query they might need while keeping it simple. 
\item "Application interest": For the authors of a specific application, it is vital that the language 
supports the operations that are requested by the application. Apart from those, all other functionality 
is rather an obstacle as it makes the language unnecessarily complex to both implement and use. 
\item "System interest": The underlying search system is responsible for efficient evaluation of queries. 
For this purpose, it is advantageous that query reformulation and optimization strategies are available and 
the language philosophy complies with the underlying data structures and algorithms. The language needs to 
support all the functionality provided by the search system.
\item "Interoperability interest": In many real-world-use scenarios it is necessary to combine information 
from several sources to get the desired knowledge. Therefore, it is desirable to have a tool that can be 
employed to query across multiple search services. A language designed for this purpose needs to be general and 
extensible.
\end{itemize}

\noindent
As we explained in the introduction, our primary objective is to create a communication interface to a 
retrieval system that is used in a number of diverse applications
and supports a wide range of search 
settings. For this purpose, the system and user points of view are most important. Interoperability is 
desirable but not critical whereas the single-application viewpoint is not relevant at all. Most of all, 
we require the language to support all the functionality enabled by MESSIF. The usability 
and optimization issues are the second most important. We are aware of the fact that language suited to these priorities will not be 
the most convenient for amateur users. However, we are more interested in providing extended functionality 
for advanced users and rely on additional software to support
beginners. Table~\ref{table:requirements} summarizes the requirements 
identified earlier and the priority levels we assign to them.

\begin{center}

\renewcommand*\arraystretch{1.3}
\begin{tabular}{|l|c|}
\hline
{\bf Language feature} & {\bf Priority} \\ \hline
\parbox{290pt}{
\vspace{3pt}
Support all standard query types 
\vspace{-1.5ex}
\begin{itemize}
 \item kNN, range, similarity joins, rKNN, skyline, distinct kNN, subsequence search, ...
 \item single- and multiple-object queries
 \item attribute-based (relational) and spacio-temporal queries
\end{itemize}
\vspace{-1ex}
} & high\\ \hline
\parbox{290pt}{
\vspace{3pt}
Allow multiple information sources and complex queries, combining attribute-based and similarity-based 
retrieval 
\vspace{3pt}} & high\\ \hline
Allow user preference settings (precise vs. approximate search, etc.) 
& high\\ \hline
\parbox{290pt}{
\vspace{3pt}
Support user-defined distance functions and distance aggregation functions
\vspace{3pt}} & high\\ \hline
Be extensible (new index structures, query types, data types) & high\\ \hline
Be user-friendly & medium\\ \hline
Be designed to allow easy query reformulation & medium\\ \hline
Provide service management tools & medium\\ \hline
Provide output presentation tools & low\\ \hline
Be compatible with MESSIF architecture & high\\ \hline
\end{tabular}
\vspace{-1ex}
\begin{table}
\caption{\label{table:requirements}Ranked list of required language features. 
}
\end{table}
\end{center}

\vspace{-3\baselineskip}

\section{Query Language Design}

The fundamental decision in a query language design resides in the choice between the construction of a brand new query language and a modification of an existing one. In this section, we discuss our choice and its impact on the architecture of retrieval systems that would implement the language.

\subsection{Overall Concept}
The desired functionality of the new language, as described in
Table~\ref{table:requirements}, comprehends the support for
standard attribute-based searching which, while not being fully sufficient anymore, still remains one of
the basic methods of data retrieval. A natural approach to creating a more powerful language therefore lies in
extending some of the existing, well-established tools for query formulation, provided that the added functionality can be nested into it. Two advantages are achieved this way: only the extended functionality
needs to be defined and implemented, and the users are not forced to learn a new syntax and semantics.

The two most frequently used formalisms for attribute data querying are the relational data model with the SQL language, and the XML-based data modeling and retrieval. As we could observe in the related work, 
both these solutions have already been employed for multimedia searching. However, there are differences in
their suitability for various use cases. The XML-based languages are well-suited for inter-system communication, but not practical for hand-typing queries because of the lengthy syntax. On the other hand,
the SQL language was designed to facilitate user-friendly data access, with the query structure imitating English sentences. In addition, SQL is backed by a strong theoretical background of relational algebra, which 
is not in conflict with content-based data retrieval and offers promising possibilities with respect to query 
optimization. Therefore, we decided to base our approach on the SQL 
language, similar to existing proposals~\cite{BarioniRTT05,BarioniRTJ09,POSTGRESQL-IE}. 

By employing the standard SQL~\cite{DBbook} we readily gain a very
complex set of functions for attribute-based retrieval but no
support at all for similarity-based searching. Since we aim at
providing a wide and extensible selection of similarity queries, it
is also not possible to employ any of the existing extensions to
SQL, which focus only on a few most common query operations.
Therefore, we created a new enrichment of both the relational data
model and the SQL syntax so that it can encompass the general
content-based retrieval as discussed in the Analysis section. The
new features will be presented in detail in the following.

In addition to attribute-based and content-based queries, some research papers distinguish a third type of retrieval -- the spacio-temporal queries. While this sort of retrieval is definitely relevant for many
applications, it does not require any functionality not available within the first two search paradigms.
We consider spatial and temporal queries to be a special instance of either attribute-based or content-based
query, depending on a particular spacio-temporal predicate: search for two time-overlapping actions would
be an instance of the former, search for time-nearest action of the latter. Naturally, specialized predicates
are needed to extract and evaluate the spacio-temporal information.

Apart from the functionality directly related to query formulation, other features mentioned in Table~\ref{table:requirements}
comprise support for query reformulation and optimization, service management, and output formatting tools. 
As for query optimization, it is not possible to create a general and extensible framework with a definite
set of optimization rules. However, we believe that the design of both the data model and operations that underlie the language itself allow to store all the necessary information that may be required by various
optimization strategies of the individual search engines. The service management will be discussed shortly in the next section in connection with extensibility issues. Output formatting is not addressed in this study
but may be easily added to the language.

\subsection{System Architecture}
In the existing proposals for multimedia query languages based on
SQL, it is always supposed that the implementing system
architecture is based on RDBMS, either directly as
in~\cite{POSTGRESQL-IE}, or with the aid of a ``blade'' interface
that filters out and processes the content-based
operations~\cite{BarioniRTJ09} while passing the regular queries to
the backing database.

Both these solutions are valid for the proposed query language.
Since we propose to extend the SQL language by adding some
language constructs, they can be easily intercepted by a ``blade'',
evaluated by an external similarity search system, and
passed back to the database where the final results are
obtained. The integration into a RDBMS follows an inverse approach.
The database SQL parser is updated to support the new language constructs
and the similarity query is evaluated by internal operators. Of
course, the actual similarity query evaluation is the corner stone
in both approaches and similarity indexes are crucial for efficient
processing.

One of our priorities is creating a user-friendly tool
for the MESSIF framework. It already supports a number of
general data types and similarity operations and is easily
extensible. The indexing algorithms can be plugged as needed
to efficiently evaluate different queries and the framework
automatically selects indexes according to the given query.
The storage backend of the MESSIF utilizes a relational database
and the functionality of the standard SQL is thus internally
supported. The data and operation model of the proposed query 
language is designed in such
a way that it is compatible with the framework.

\section{Query Language Specification}

In this section we present the \QLname{}, an extension of the SQL
query language which supports advanced multimedia searching in a
flexible and extensible way. The language can be used as a
communication interface to any retrieval system that complies with
the abstract data model and operations described in
Section~\ref{sec:data-model-operations} and is able
to parse and process the SQL syntax with the enrichment introduced
in Section~\ref{sec:syntax-semantics}. In the end of the section we shortly discuss the
extensibility of our design and the query processing procedure.

\subsection{Data Model and Operations}
\label{sec:data-model-operations}

The core of any information management system is formed by data
structures that store the information, and operations that allow to
access and change it. To provide support for the content-based
retrieval, we need to revisit the data model employed by the
standard SQL and adjust it to the needs of complex data management. 

It is important to clarify here that we do not aim at defining a
sophisticated algebra for content-based searching, which is being
studied elsewhere. For the purpose of the query language, we only
need to establish the basic building blocks. Our model is in fact a
simplified version of the general framework presented
in~\cite{Subrahmanian98}. Contrary to the theoretical algebra
works, we do not study the individual operations and their
properties but let these be defined explicitly by the underlying
search systems. However, we introduce a more fine-grained
classification of objects and operations to enable their easy
integration into the query language.

\subsubsection{Data Model}

On the concept level, multimedia objects can be analyzed using
standard entity-relationship (ER) modeling.  In the ER terminology,
a real-world object is represented by an {\em entity}, which is
formed by a set of descriptive object properties -- {\em
attributes}. The attributes need to contain all information
required by target applications. In contrast to common data types
used in ER modeling, which comprise mainly text and numbers,
attributes describing multimedia objects are often of more complex
types, such as image or sound data, time series, etc.  The actual
attribute values form an {\em n-tuple} and a set of n-tuples of the
same type constitute a {\em relation}.

Relations and attributes (as we shall continue to call the elements
of n-tuples) are the basic building blocks 
of the Codd's relational data model and algebra~\cite{Codd90}, upon which the SQL language is based. This 
model can also be employed for complex data retrieval but we need to introduce some extensions. A relation is 
traditionally defined as a subset of the Cartesian product of sets $D_1$ to $D_n$, $D_i$ being the {\em 
domain} of attribute $A_i$. The standard operations over relations (selection, projection, etc.) are defined 
using first-order predicate logic and can be readily applied on any data, provided the predicates can be 
reasonably evaluated over the data. To control this, we use the concept of {\em data type} that 
encapsulates both a specification of an attribute domain and the functions that can be applied on members of 
this domain. Let us note here that Codd used a similar concept of {\em extended data type} in~\cite{Codd90}, 
however he only worked with several special properties of the data type, in particular the total ordering. As 
we shall discuss presently, our approach is much more general. We allow for an infinite number of data types, 
as opposed to the traditional finite set of types that appear in most data management systems. The 
individual data types directly represent the objects (e.g. text, image, video, sound), or some derived 
information (e.g. color histogram vector). The translation of one
data type into another can be realized by so called
{\em extractors}, a special type of functions defined for each data
type.

\begin{figure}[ht]
\vspace{-0.7em}
\centerline{
\includegraphics[width=0.9\textwidth]{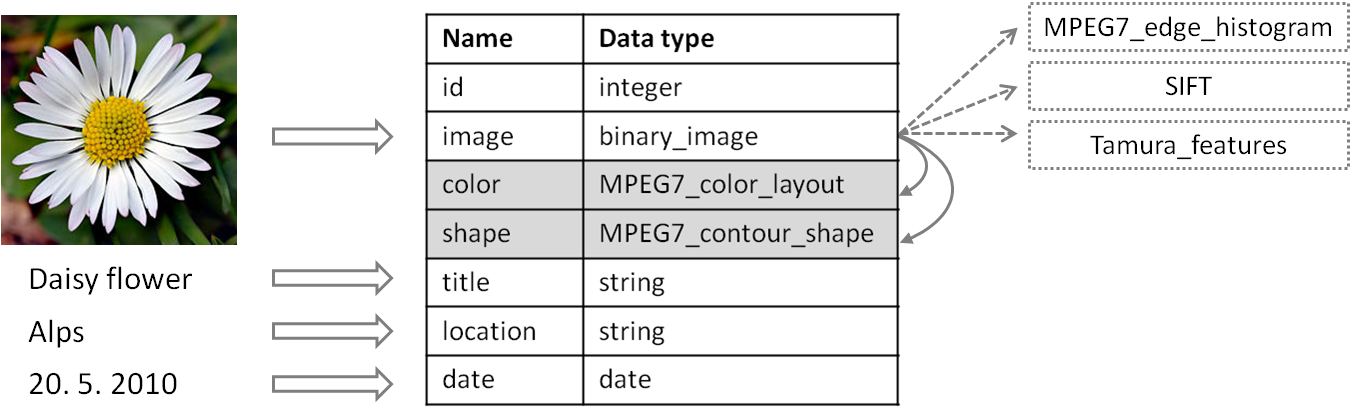}
}
\caption{\label{fig:object-relation}Transformation of image object
into a relation. Full and dashed arrows on the 
right side depict materialized and available data type extractors, respectively. 
}
 \vspace{-2em}
\end{figure}

According to the best-practices of data modeling~\cite{DBbook}, redundant data should not be present in the 
relations, which also concerns derived attributes. The rationale is that the derived information only requires
extra storage space and introduces the threat of data inconsistency. Therefore, the derived attributes should 
only be computed when needed in the process of data management. In case of complex data, however, the 
computation (i.e. the extraction of derived data type) can be very costly. Thus, it is more suitable to allow 
storing some derived attributes in relations, especially when these are used for data indexing. Naturally, 
more extractors may be available to derive additional attributes
when asked for. Figure~\ref{fig:object-relation} depicts 
a possible representation of an image object in a relation.

\subsubsection{Operations on Data Types}

As we already stated, each data type consists of a specification of a domain of values, and a listing of 
available functions. As some of the functions are vital for the formulation and execution of the algebra 
operations, we introduce several special classes of functions that may be associated with each data type. 

\begin{itemize}
\item {\em Comparison functions}: Functions of this type define total ordering of the domain 
($f_C:D \times D \rightarrow \{<, =, >\}$). When a comparison 
function is available, standard indexing methods such as B-trees can be applied and queries using value 
comparison can be evaluated. Comparison functions are typically not available for multimedia data types and 
the data types derived from them, where no meaningful ordering of values can be defined. 
\item {\em Distance functions}: In the context of datatypes we focus on basic distance functions that evaluate 
the dissimilarity between two values from a given data domain ($f_D:D \times D \rightarrow \mathbb{R}_{0}^+$).
The zero distance represents the maximum possible similarity -- identity. 
We do not impose any additional restrictions on the behavior of $f_D$ in general, but there exists a way of 
registering special properties of individual functions that will be discussed later.
More than one distance function can be assigned to a data
type, in that case one of the functions needs to be denoted as default. When more distance functions are 
available for a given data type, a specification of the preferred distance can be part of relation definition.
In case no distance function is provided, a trivial {\em identity distance} is associated to the data type, 
which assigns distance 0 to a pair of identical values and distance $\infty$ to any other input. 
\item {\em Extractors}: Extractor functions transform values of one data type into the values of a different
data type ($f_E:D_i  \rightarrow D_j$). Extractors are typically used on complex unstructured data types (such
as binary image) to produce data types more suitable for indexing and retrieval (e.g. color descriptor). An
arbitrary number of extractors can be associated to each data type.  
\end{itemize}  

\noindent
In addition to the declaration of functionality, each of the mentioned operations can be equipped by 
a specification of various properties. The list of properties that are considered worthwhile is inherent 
to a particular retrieval system and depends on the data management tools employed. For instance, many 
indexing and retrieval techniques for similarity searching rely on certain properties of distance functions, 
such as the metric postulates or monotonicity. To be able to use such a technique, the system needs to ascertain 
that the distance function under consideration satisfies these requirements. To solve this type of inquiries in 
general, the set of properties that may influence the query processing is defined, and the individual 
functions can provide values for those properties that are relevant for the particular function. To continue with 
our example, the Euclidean distance will declare that it satisfies the metric postulates as well as monotonicity, 
while the MinimumValue distance only satisfies monotonicity. Another property worth registering is a lower-bounding relationship between two distance functions, which may be utilized during query evaluation.

\subsubsection{Operations on Relations}

The functionality of a search system is provided by the operations
that can be evaluated over relations. In addition to standard
selection and join operations, multimedia search engines need to
provide operations for various types of similarity-based retrieval.
Due to the diversity of possible approaches to searching, we do not
introduce a fixed set of operations that need to be available in a
search system, but expect each system to maintain its own list of
operations. Each operation needs to specify its input, which
consists of 
1) number of input relations (one for simple queries, multiple for joins), 
2) expected query objects (zero, singleton, or arbitrary set), 
3) arbitrary number of operation-specific parameters, which may
typically contain a specification of a distance function, distance
threshold, or query operation execution parameters such as
approximation settings.
Apart from a special case discussed later the operations return
relations, typically with the scheme of the input relation or the
Cartesian product of input relations. In case of similarity-based
operations the scheme is enriched with additional {\em distance}
attribute which carries the information about the actual distance
of a given result object with respect to the distance function employed
by the search operation.

Similar to operations on data types, operations on relations may also exhibit special properties that can 
be utilized with advantage by the retrieval system. In case of data retrieval operations, the properties
are mainly related to query optimization. As debated earlier, it is not possible to define general optimization
rules for a model with a variable set of operations. However, a particular retrieval system can maintain
its own set of optimization rules together with the list of operations. 

A special subset of operations on relations is formed by functions that produce scalar values. Among these,
the most important are the {\em generalized distance} functions that 
operate on relations and return a single number, representing the distance of objects more complex than 
values from a given attribute domain. The input of these functions contains 1) a relation representing the
object for which the distance needs to be evaluated, 2) a relation with one or more query objects, and 3)
additional parameters when needed. Similar to basic distance functions, generalized distance functions need to be 
treated in a special way since their properties often significantly influence the processing of a query. 
Depending on the architecture of the underlying search engine it may be beneficial to distinguish more 
types of generalized distance functions. For the MESSIF architecture in particular, we define the following
two types:
\begin{itemize}
\item {\em Set distance $f_{SD}: 2^D \times D \times (D \times D \rightarrow \mathbb{R}_{0}^+) \rightarrow 
\mathbb{R}_{0}^+$}: The set distance function allows to evaluate the similarity of object to a set of query objects 
of the same type, employing the distance function defined over the respective object type. In a typical implementation, 
such function may return the minimum of the distances to individual query objects.
\item {\em Aggregated distance $f_{AD}: (D_1 \times ... \times D_n) \times (D_1 \times ... \times D_n) 
\times ((D_1 \times D_1 \rightarrow \mathbb{R}_{0}^+) \times ... \times (D_n \times D_n \rightarrow 
\mathbb{R}_{0}^+)) \rightarrow \mathbb{R}_{0}^+$}: The aggregation of distances is frequently employed 
to obtain a more complex view on object similarity. For instance, the similarity of images can be evaluated 
as a weighted sum of color- and shape-induced similarities. The respective weights of the partial 
similarities can be either fixed, or chosen by user for a specific query. Though we do not include 
the user-defined parameters into the definitions of the distances for easier readability, these are 
naturally allowed in all functions.  
\end{itemize}   
 
\subsubsection{Data Indexing}

While not directly related to the data model, data indexing methods are a crucial component of a retrieval 
system. The applicability of individual indexing techniques is limited by the properties of the target data. 
To be able to control the data-index compatibility or automatically choose a suitable index, the search 
system needs to maintain a list of available indices and their properties. The properties can then be verified 
against the definition of the given data type or distance function (basic or generalized). Thus, metric index structures for 
similarity-based retrieval can only be made available for data with metric distance function, whereas 
traditional B-trees may be utilized for data domains with total ordering. It is also necessary to specify which
search operations can be supported by a given query, as different data processing is needed e.g. for the 
nearest-neighbor and reverse-nearest-neighbor queries. Apart from the specialized indices, any search system 
inherently provides the basic {\em Sequential Scan} algorithm as a default data access method that can support
any search operation.   

\subsection{\QLname{} Syntax and Semantics}
\label{sec:syntax-semantics}

The \QLname{} language is designed to provide a user-friendly
interface to state-of-the-art multimedia search systems. Its main
contribution lies in enriching the standard SQL by new language
constructs that enable to issue all kinds of content-based queries
in a standardized manner. In accordance with the declarative
paradigm of SQL, the new language constructs allow to describe the
desired results while shielding users from the execution issues. On
the syntactical level, the \QLname{} contributes mainly to the query
formulation tools of SQL. Data modification and control commands
are not discussed in this paper since their adaptation to the
generalized data types and operations is straightforward. On the
semantic level, however, the original SQL is significantly enriched
by the introduction of the unlimited set of complex data types and
operations over them.
 
A \QLname{} query statement follows the same structure as standard
SQL, being composed of the six basic clauses SELECT, FROM, WHERE,
GROUP BY, HAVING, and ORDER BY, with their traditional
semantics~\cite{DBbook}. The extended functionality is mainly
provided by a new construct called SIMSEARCH, which is embedded
into the FROM clause and allows to search by similarity, combine
multiple sources of information, and reflect user preferences.
Prior to a detailed description of the new primitives, we present
the overall query syntax with the SIMSEARCH construct in the
following scheme:

\renewcommand*\arraystretch{1.2}

\vspace{0.5\baselineskip}
\begin{center}
\noindent
\begin{tabular}{l l}
{\bf SELECT} & [TOP n $\vert$ ALL] \\
& \{attribute $\vert$ ds.distance $\vert$ ds.rank $\vert$ f(params)\} [, ...] \\
{\bf FROM} & \{dataset $\vert$ \\
& SIMSEARCH [:obj [, ...]] \\
& \hskip 20pt IN data\_source AS ds [, data\_source2 [, ...]]\\
& \hskip 20pt BY \{attribute [DISTANCE FUNCTION \\
& \hskip 90pt distance\_function(params)] \\
& \hskip 40pt  $\vert$ distance\_function(params)\}\\
& \hskip 20pt  [METHOD method(params)] \\
{\bf WHERE} & /* restrictions of attribute values */ \\
{\bf ORDER BY} & \{attribute $\vert$ ds.distance [, ...]\}
\end{tabular}
\end{center}
\vspace{0.5\baselineskip}

\noindent
In general, there are two possible approaches to incorporating primitives for content-based retrieval into the SQL syntax. We can either make the similarity search results form a new information resource on the level of 
other data collections in the FROM clause (an approach used
in~\cite{POSTGRESQL-IE}), or handle the similarity as another of
the conditions applied on candidate objects in the WHERE clause
(exercised in~\cite{BarioniRTJ09,MOQL,Amato97,GaoWWP04}). However, the latter approach requires standardized predicates for various types of similarity queries, their parameters etc., which is difficult to achieve in case an extensible set of search operations and algorithms is to be supported. In addition, the similarity predicates are of different nature than attribute-based predicates and their efficient evaluation requires specialized data structures. 
Therefore, we prefer to handle similarity-based retrieval as an independent information source. For this, we 
only standardize the basic structure and expected output, which can be implemented by any number of search methods of the particular search engine. 

As anticipated, the similarity-based retrieval is wrapped into the SIMSEARCH language construct, which produces
a standard relation and can be seamlessly integrated into the FROM clause. The SIMSEARCH expression is composed
 of the following parts:
\begin{itemize}
\item Specification of query objects: The selection of query objects follows immediately after the SIMSEARCH 
keyword. An arbitrary number of query objects can be issued, each object being in fact an attribute that can 
be compared to attributes of the target relations. The query object (attribute) can be represented directly 
by the attribute value, by a reference to object provided by an application, or by a nested query that produces 
the query object(s). The query objects need to be type-compatible
with the attributes of target relation they are to be compared to.
Often the extractor functions can be used with advantage on the
query objects.
\item Specification of a target relation: The keyword IN introduces the specification of one or more 
relations, elements of which are processed by the search algorithm. Naturally, each relation can be 
produced by a nested query. 
\item Specification of a distance function: An essential part of a content-based query is the specification
of a distance function. The BY subclause offers three ways of defining the distance: calling a distance 
function associated to an attribute, referring directly to a distance function provided by the search engine, 
or constructing the function within the query. In the first case, it is sufficient to enter the name of 
attribute to invoke its default distance function. Non-default distance function of an attribute needs to
be selected via the DISTANCE FUNCTION primitive that also allows to
pass additional parameters for the distance function if necessary.
The last case allows greater freedom of specifying the distance
function by user, but both the attributes for which the distance is
to be measured must be specified. A special function DISTANCE$(x,y)$
can be used to call the default distance function defined for the
given data type of attributes $x,y$. The nuances of referring to a distance function can be observed in the following:\\
\vspace{0.1em}\\
\renewcommand*\arraystretch{1.0}
\begin{tabular}{l}
\hskip 20 pt SIMSEARCH ... BY color \\
\hskip 40 pt /* search by default distance function of the color attribute */\\
\hskip 20 pt SIMSEARCH ... BY color DISTANCE FUNCTION color\_distance \\
\hskip 40 pt /* search by {\em color\_distance} function of the color attribute */\\
\end{tabular}

\vspace{1em}
\begin{tabular}{l}
\hskip 20 pt SIMSEARCH ... BY some\_special\_distance(qo, color, params)\\ 
\hskip 40 pt /* search by {\em some\_special\_distance} applied to
query\\
\hskip 55 pt    object {\em qo}, color attribute, and additional parameters */\\
\hskip 20 pt SIMSEARCH ... BY DISTANCE(qoc, color)+DISTANCE(qos, shape)\\ 
\hskip 40 pt /* search by a user-defined sum the of the default distance \\
\hskip 55 pt    functions on color and shape attributes */\\
\end{tabular}
\vspace{1em}

\item Specification of a search method: The final part of the SIMSEARCH construct specifies the search methods or, in other words, the query type. Users may choose from the list of methods offered by the search system.
It can be reasonably expected that every system supports the basic nearest neighbor query, therefore this is
considered a default method in case no other is specified with the METHOD keyword. The default nearest neighbor search returns all n-tuples from the target relation unless the number of nearest neighbors is specified in
the SELECT clause by the TOP keyword.
\end{itemize}

\noindent
The complete SIMSEARCH phrase returns a relation with a scheme of
the target relation specified by the IN keyword, or the Cartesian
product in case of more source relations. Moreover, information about distance of each n-tuple of the result set computed during the content-based retrieval is available. This can be used in other
clauses of the query, referenced either as DISTANCE, when only one
distance evaluation was employed, or prefixed with the named data
source in the clause when ambiguity should arise (e.g. ds.DISTANCE).

\subsection{Extensibility}

The extensibility of the \QLname{} language relies on the
possibility to define a set of data types, functions, query
operations, and index structures supported by each retrieval
engine. The information about the system functionality should be
maintained in special relations with standardized structure, which
would allow automatic service discovery. The design of these
relations will be subject of our future work.

\subsection{Query Processing}

The query processing is a complex procedure that needs to be
designed carefully with respect to the architecture of a given
retrieval system. Nonetheless, the following succession of basic
steps will always form the basic structure of the processing. Fist
of all, a parser identifies the individual objects and operations
contained in the query expression. Using registered properties, the
query processing unit checks the compatibility. When successful, an
evaluation plan is composed. For its construction, the system may
use the available indices together with the registered properties
of attributes, indices, and functions. The optimal evaluation plan
is eventually executed and the results returned to the user.

\section{Example Scenarios}

To illustrate the wide applicability of the \QLname{} language, we now present several query examples for various use-case scenarios found in image and video retrieval. Each of them is accompanied by a short comment on the interesting language features employed. For the examples, let us suppose that the following set of relations, data types and functions is available in the retrieval system:

\renewcommand*\arraystretch{1.0}
\renewcommand{\tabcolsep}{0.2cm}

\begin{itemize}

\item {\bf image} relation: register of images\\
\begin{center}
\vspace{-0.7em}
\begin{tabular}{|l |l| l|}
\hline
id & integer & identity\_distance {\em (default)}\\ \hline
image & binary\_image & identity\_distance {\em (default)}\\ \hline
color & number\_vector & mpeg7\_color\_layout\_metric {\em (default)}\\ 
 & & L1\_metric \\ \hline
shape & number\_vector & mpeg7\_contour\_shape\_metric {\em (default)}\\ 
 & & L2\_metric \\ \hline
title & string & tf\_idf {\em (default)}\\ \hline
location & string & simple\_edit\_distance {\em (default)}\\ \hline
date & date & L1\_metric {\em (default)}\\ \hline
\end{tabular}
\end{center}
\vspace{0.5em}

\item {\bf video\_frame} relation: list of video frames\\
\begin{center}
\vspace{-0.7em}
\begin{tabular}{|l |l| l|}
\hline
id & integer & identity\_distance {\em (default)}\\ \hline
video\_id & integer & identity\_distance {\em (default)}\\ \hline
video & binary\_video & identity\_distance {\em (default)}\\ \hline
face\_descriptor & number\_vector & mpeg7\_face\_metric {\em (default)}\\ \hline
subtitles & string & tf\_idf {\em (default)}\\ \hline
time\_second & long & L1\_metric {\em (default)}\\ \hline
\end{tabular}
\end{center}
\vspace{0.5em}

\item {\bf keyword} relation: a simple table of keywords which can be related to an image/video (e.g. web gallery tags)\\
\begin{center}
\vspace{-0.7em}
\begin{tabular}{|l |l| l|}
\hline
id & integer & identity\_distance {\em (default)}\\ \hline
value & string & simple\_edit\_distance {\em (default)}\\
&& weighted\_edit\_distance \\ \hline
\end{tabular}
\end{center}
\vspace{0.5em}
 
\item {\bf image\_keyword} relation: keywords associated with an image\\
\begin{center}
\vspace{-0.7em}
\begin{tabular}{|l |l| l|}
\hline
image\_id & integer & identity\_distance {\em (default)}\\ \hline
keyword\_id & integer & identity\_distance {\em (default)}\\ \hline
\end{tabular}
\end{center}
\vspace{0.5em}

\end{itemize}

\renewcommand{\tabcolsep}{0.1cm}

\subsubsection{Query 1} {\em Retrieve 30 most similar images to a given example}

\begin{center}
\noindent
\begin{tabular}{l l}
SELECT & TOP 30 id, distance \\
FROM   & SIMSEARCH :queryImage IN image BY shape \\ 
\end{tabular}
\end{center}

\noindent
This example presents the simplest possible similarity query. It
employs the default nearest neighbor operation over the shape
descriptor with its default distance function. User does not need
any knowledge about the operations employed, only selects the means
of similarity evaluation. The supplied parameter $queryImage$
represents the MPEG7\_contour\_shape type of a query image
(provided by surrounding application). The output of the search is
the list of identifiers of the most similar images as well as the
distance measured between the query image and the respective image
in the database.

\subsubsection{Query 2} {\em Retrieve all variants of the word 'feather' with maximally two typos}

\begin{center}
\noindent
\begin{tabular}{l l}
SELECT & value \\
FROM   & SIMSEARCH 'feather' IN keyword BY value\\
& \hskip 20pt  DISTANCE FUNCTION weighted\_edit\_distance(1,2,2) \\ 
WHERE & distance $<=$ 2
\end{tabular}
\end{center}

\noindent
This time, a simple range query is required. The user selected a non-default distance function for the evaluation of similarity, which is adjusted by user-defined weights for the edit, insert and remove actions. Depending on the underlying search system, the query can be either reformulated and evaluated as a range query, or evaluated as a nearest neighbor query with subsequent result object filtering.

\subsubsection{Query 3} {\em Find all pairs of keywords that are within edit distance 1 (we may suppose that these are candidates for typos)}

\begin{center}
\noindent
\begin{tabular}{l l}
SELECT & * \\
FROM   & SIMSEARCH \\
& \hskip 20pt IN keyword AS k1, keyword AS k2 \\
& \hskip 20pt BY simple\_edit\_distance(k1.value, k2.value) \\
& \hskip 20pt METHOD MessifSimilarityJoin(1) \\ 
\end{tabular}
\end{center}

\noindent
In this case, a similarity join with a threshold value 1 is
required. The similarity join needs no query objects, is defined
over two relations, and requires explicit reference to a distance
function with the input parameters.

\subsubsection{Query 4} {\em Retrieve images most similar to a set
of examples (e.g. identifying a flower by supplying several photos)}

\begin{center}
\noindent
\begin{tabular}{l l}
SELECT & TOP 1 title \\
FROM   & SIMSEARCH \\
& \hskip 40pt extract\_MPEG7\_color\_layout(:o1) AS co1,\\ 
& \hskip 40pt extract\_MPEG7\_color\_layout(:o2) AS co2,\\
& \hskip 40pt extract\_MPEG7\_contour\_shape(:o3) AS sh3\\
& \hskip 20pt IN image \\
& \hskip 20pt BY minimum(DISTANCE(co1, color), DISTANCE(co2, color), \\
& \hskip 40pt DISTANCE(sh3,shape)) \\ 
\end{tabular}
\end{center}

\noindent
This query represents an example of a multi-object query, input of
which are binary images that are transformed to the required
descriptors via extractors. Alternatively, the query objects could
be provided as a result of a nested query. The minimum aggregation
function employed for similarity evaluation would be formally
defined on attributes and their respective distance functions. Here
it is applied on the distances to individual objects only, as these
are internally linked to the individual attributes and distance
functions. Note that the default distance functions of the
respective attributes are applied using DISTANCE$(x,y)$ construct.

\subsubsection{Query 5} {\em Retrieve all videos where Obama and Bush appear}

\begin{center}
\noindent
\begin{tabular}{l l}
SELECT & DISTINCT vf1.video\_id \\
FROM   & SIMSEARCH :ObamaFace IN video\_frame AS vf1 BY face\_descriptor \\ 
      & \hskip 20pt METHOD rangeQuery(0.01)  \\
& INNER JOIN \\
& SIMSEARCH :BushFace IN video\_frame AS vf2 BY face\_descriptor \\
      & \hskip 20pt METHOD rangeQuery(0.01)\\
& ON (vf1.video\_id = vf2.video\_id) \\
\end{tabular}
\end{center}

\noindent
This query employs a join of two similarity search results, each of
which uses a range query operation to retrieve objects very similar
to the given example.
   
\subsubsection{Query 6} {\em Retrieve all videos where the Vesuvius mountain appears (image similarity) and
   a commentator mentions volcanoes (speech/text similarity) within two minutes (time aggregation)}

\begin{center}
\noindent
\begin{tabular}{l l}
SELECT & vf1.video\_id \\
FROM  & SIMSEARCH IN\\ 
 & \hskip 20pt SIMSEARCH \\
 & \hskip 40pt extract\_MPEG7\_color\_layout(:VesuvImage) AS co,\\ 
& \hskip 40pt extract\_MPEG7\_contour\_shape(:VesuvImage) AS sh\\
 & \hskip 20pt IN video\_frame AS vf1 \\
& \hskip 40pt BY weight\_sum((DISTANCE(shape,sh), 0.7), \\
& \hskip 110pt (DISTANCE(color, co), 0.2))\\
& \hskip 40pt METHOD MessifRangeQuery(0.1,15000)\\  
&  \hskip 20pt     NATURAL JOIN\\
&	 \hskip 20pt    SIMSEARCH 'vulcano' IN video\_frame AS vf2 \\
& \hskip 40pt BY subtitles \\
& \hskip 40pt METHOD MessifRangeQuery(0.1,15000)\\
&		   BY DISTANCE(vf1.time\_second, vf2.time\_second) \\
&          AS sim\_frames \\
WHERE & sim\_frames.distance $<=$ 120 
\end{tabular}
\end{center}

\noindent
In this example, multiple modalities are combined to produce the result. In addition, the user selected
a special search method that enables to set approximation in work (the second parameter of the search method
is the maximum number of objects that may be visited in query processing).

\subsubsection{Query 7} {\em Retrieve text annotation for a given unknown image, exploiting the keywords 
associated to similar images}

\begin{center}
\noindent
\begin{tabular}{l l l l}
SELECT & \multicolumn{3}{l}{value} \\ 
FROM   & \multicolumn{3}{l}{keyword} \\ 
WHERE  & \multicolumn{3}{l}{id IN (} \\
&	         \hskip 20pt SELECT & \multicolumn{2}{l}{TOP 10 keyword\_id, count(image\_id) AS frequency} \\ 
&          \hskip 20pt FROM & \multicolumn{2}{l}{(} \\
&&		            \hskip 20pt SELECT &  TOP 30 id\\ 
&&                \hskip 20pt FROM   &  SIMSEARCH :Image IN image \\
&&&                     \hskip 20pt BY color DISTANCE\\
&&&                     \hskip 20pt FUNCTION L1\_metric \\
&&          \multicolumn{2}{l}{) AS simimage} \\
&&           \multicolumn{2}{l}{INNER JOIN image\_keyword} \\
&&           \multicolumn{2}{l}{\hskip 20pt ON (simimage.id = image\_keyword.image\_id)}\\
&	         \hskip 20pt GROUP BY & \multicolumn{2}{l}{keyword\_id} \\
&	         \hskip 20pt HAVING &  \multicolumn{2}{l}{frequency $>$ 3} \\
&	         \hskip 20pt ORDER BY & \multicolumn{2}{l}{frequency DESC} \\
&           ) & & \\
\end{tabular}
\end{center}

\noindent
The final example presents a content-based search nested into a
complex expression of the traditional SQL.

\section{Conclusions and Future Work}
The two main contributions of this paper are the analysis of
requirements for a query language, and the proposal of a query
language for retrieval over complex data domains. The presented
language is backed by a general model of data structures and
operations, which is applicable to a wide range of search systems
that offer different types of content-based functionality.
Moreover, the support for data indexing and query optimization is
inherently contained in the model. The \QLname{} language extends the
standard SQL by new primitives that allow to formulate
content-based queries in a flexible way, taking into account the
functionality offered by a particular search engine.

The proposal of the language was influenced by the MESSIF framework
that offers the functionality of executing complex similarity
queries on arbitrary index structures but lacks a user-friendly interface
for advanced querying. Having laid the formal foundations of the query interface
here, we will proceed with the implementation of a language parser which will translate the query into MESSIF for the actual evaluation.

In the future, we plan to research the possibilities of adapting
the existing optimization strategies to utilize the reformulation
capabilities of the proposed extension. Furthermore, we would like
to create an intuitive (graphical) query formulation tool and,
possibly, a conversion mechanism into the MPEG7 Query Format for
inter-system communication. Also, the syntax for the various service discovery tools needs to be established.

\vspace{-1ex}
\section*{Acknowledgments}
\vspace{-1ex}
{\small 
This work has been partially supported by Brno PhD Talent Financial Aid and by the national research projects GAP 103/10/0886 and VF 20102014004.
}

\bibliographystyle{splncs03}

\end{document}